\documentclass[sigconf]{acmart}

\copyrightyear{2026}
\acmYear{2026}
\setcopyright{cc}
\setcctype{by}
\acmConference[ICSE-NIER '26]{2026 IEEE/ACM 48th International Conference on Software Engineering}{April 12--18, 2026}{Rio de Janeiro, Brazil}
\acmBooktitle{2026 IEEE/ACM 48th International Conference on Software Engineering (ICSE-NIER '26), April 12--18, 2026, Rio de Janeiro, Brazil}
\acmPrice{}
\acmDOI{10.1145/3786582.3786845}
\acmISBN{979-8-4007-2425-1/2026/04}

\AtBeginDocument{%
  \providecommand\BibTeX{{%
    \normalfont B\kern-0.5em{\scshape i\kern-0.25em b}\kern-0.8em\TeX}}}

\usepackage[linesnumbered,ruled,vlined]{algorithm2e}
\usepackage{textcomp}
\usepackage{xcolor}
\usepackage{subcaption}
\usepackage{multirow}
\usepackage{makecell}
\usepackage{enumitem}
\usepackage{listings}

\begin{document}

\newcommand{\todo}[1]{\textbf{\color{red}{Ravishka: #1} }}

\newcommand{\wei}[1]{\textcolor{blue}{Wei: [#1]}}

\newcommand{\ww}[1]{\textcolor{blue}{Weihang: [#1]}}

\newcommand{\td}[1]{\textcolor{blue}{ToDo: [#1]}}

\newcommand{\zijie}[1]{{\color{purple}[Zijie: #1]}}

\newcommand{\zeqing}[1]{{\textcolor{blue}[Zeqing: #1]}}

\newcommand{\jiajun}[1]{{\color{cyan}[Jiang: #1]}}

\newcommand{\distance}{4pt}
\setlength{\textfloatsep}{\distance}

\newcommand\mycommfont[1]{\small\ttfamily\textcolor{violet}{#1}}
\SetCommentSty{mycommfont}

\lstdefinestyle{Cpp}{ 
	language=C++,
	basicstyle=\scriptsize\ttfamily, 
	breakatwhitespace=false, 
	breaklines=true, 
	captionpos=b, 
	commentstyle=\color[rgb]{0.0, 0.5, 0.69},
	deletekeywords={}, 
	escapeinside={<@}{@>},
	firstnumber=1, 
	frame=lines, 
	frameround=tttt, 
	keywordstyle={[1]\color{blue!90!black}},
	keywordstyle={[3]\color{red!80!orange}},
	morekeywords={String,int}, 
	numbers=left, 
	numbersep=-8pt, 
	numberstyle=\tiny\color[rgb]{0.1,0.1,0.1}, 
	rulecolor=\color{black}, 
	showstringspaces=false, 
	showtabs=false, 
	stepnumber=1, 
	stringstyle=\color[rgb]{0.58,0,0.82},
	tabsize=2, 
	backgroundcolor=\color{white}
}

\title{When to Answer and When to Defer: A Decision Framework for Reliable Code Predictions}

\author{Ravishka Rathnasuriya}
\orcid{0009-0005-6129-2865}
\affiliation{%
  \institution{The University of Texas at Dallas
}
  \country{USA}
}
\email{ravishka.rathnasuriya@utdallas.edu}

\author{Wei Yang}
\orcid{0000-0002-5338-7347}
\affiliation{%
  \institution{The University of Texas at Dallas}
  \country{USA}
}
\email{wei.yang@utdallas.edu}

\begin{abstract}

Code language models are increasingly adopted for both understanding and generative tasks. Despite their success, these models frequently produce overconfident incorrect predictions and underconfident correct predictions, undermining their reliability in deployment. Practical deployment demands three capabilities: accurately estimating the likelihood of correctness, abstaining on uncertain predictions, and invoking external mechanisms to validate or repair abstained outputs.

Existing calibration and uncertainty estimation methods, primarily developed for natural language tasks, do not readily transfer to code. Notably, post-hoc calibration techniques often reduce probability misalignment but fail to improve the ranking of predictions by correctness likelihood—a requirement for selective prediction under partial coverage. Furthermore, most approaches treat uncertainty as a passive indicator rather than an actionable signal.

This work introduces a unified framework that integrates uncertainty estimation, model calibration, and tool-based abstention handling for code models. The proposed design enables models to assign reliable correctness probabilities, abstain under uncertainty, and invoke lightweight program analysis procedures to process abstained cases. By combining these components within a single deployment-oriented workflow, this framework supports risk-aware, coverage-controlled use of code models across both classification and generation settings.

\end{abstract}
\maketitle

\section{Introduction}

Code language models are increasingly applied across two primary classes of tasks: classification tasks, including vulnerability detection and defect prediction~\cite{tian2023fly, naturalattack, Zhang2023Challenging, lu2021codexglue, yefet2020adversarial,hu2023codes,li2021estimating,van2020tailoring,peng2018t}, and generative tasks, such as code completion, synthesis, and repair~\cite{roziere2023code,guo2024deepseek,jiang2024self,evalplus,evalperf}. These models are now integral to developer workflows via integration into IDEs, CI pipelines, and automated programming assistants. However, despite their success, these models frequently exhibit poor alignment between predictive confidence and actual correctness~\cite{spiess2024calibration,zhou2024calibration}. In practice, they often assign high confidence to incorrect predictions and low confidence to correct ones which undermine their reliability, especially when used in high-stakes or partially automated environments.

Reliable deployment of code models requires more than producing high-confidence outputs. It requires mechanisms that can estimate, for each prediction, whether it is likely to be correct. This entails three key capabilities. First, models must produce a well-calibrated estimate of per-sample correctness, that is, the predicted confidence should reflect the true likelihood of correctness. Second, systems must support selective abstention, allowing models to withhold predictions when uncertainty is high. Third, there must be principled mechanisms for acting on abstained inputs, for example, by routing them to lightweight analysis tools, heuristic checks, or human review. These capabilities enable a deployment setting where model outputs can be accepted only when trustworthy, and otherwise deferred for further handling. 

While uncertainty estimation and calibration have been studied extensively in other domains~\cite{guo2017calibration,zhou2024calibration,vasilev2023calibration, gupta2020calibration}, existing methods face fundamental limitations when applied to code. Confidence signals often fail to capture the structured and semantic nature of code predictions. Similarly, post-hoc calibration techniques~\cite{guo2017calibration, zhou2024calibration,spiess2024calibration} may reduce average miscalibration but do not improve per-sample reliability in a way that supports effective abstention decisions. Moreover, these methods typically operate in isolation, with no integrated support for resolving the uncertain cases they identify.

Empirical studies from our prior work further underscore this gap~\cite{spiess2024calibration,zhou2024calibration}. We find that uncertainty signals~\cite{hendrycks2018baseline,steinhardt2016unsupervised,shannon1948mathematical,monarch2021human,gal2016dropout,lakshminarayanan2017simple,rathnasuriya2025codeimprove,rathnasuriya2025framework,rathnasuriya2025fly,vashurin2024benchmarking}, across both classification and generation tasks, often correlate poorly with correctness, particularly in code scenarios involving rare patterns, ambiguous syntax, or long-range dependencies. Calibration improves score alignment but fails to improve abstention efficacy: the calibrated probabilities do not reliably distinguish between trustworthy and untrustworthy predictions. These findings motivate the need for a unified formulation that not only estimates uncertainty and calibrates output confidence, but also makes abstention operationally useful.

We propose a deployment-oriented framework for code models that addresses this gap. Our formulation supports both classification and generation tasks, and integrates three essential components: (i) the construction of more informative uncertainty signals tailored to code, (ii) both pre-deployment and post-deployment calibration strategies that improve the interpretability of model confidence, and (iii) the incorporation of model–context–protocol (MCP)~\cite{sarkar2025survey} tools to handle abstained predictions through lightweight rule-based, static, or semantic analyses. This approach supports coverage-controlled decision-making and enables actionable abstention pipelines without requiring model retraining as a prerequisite. Together, these components provide a foundation for robust and risk-aware deployment of code models in real-world software engineering workflows.

\section{Our Envision}
\label{problem}

\begin{figure}[!htbp]
\centering
\includegraphics[width=\columnwidth]{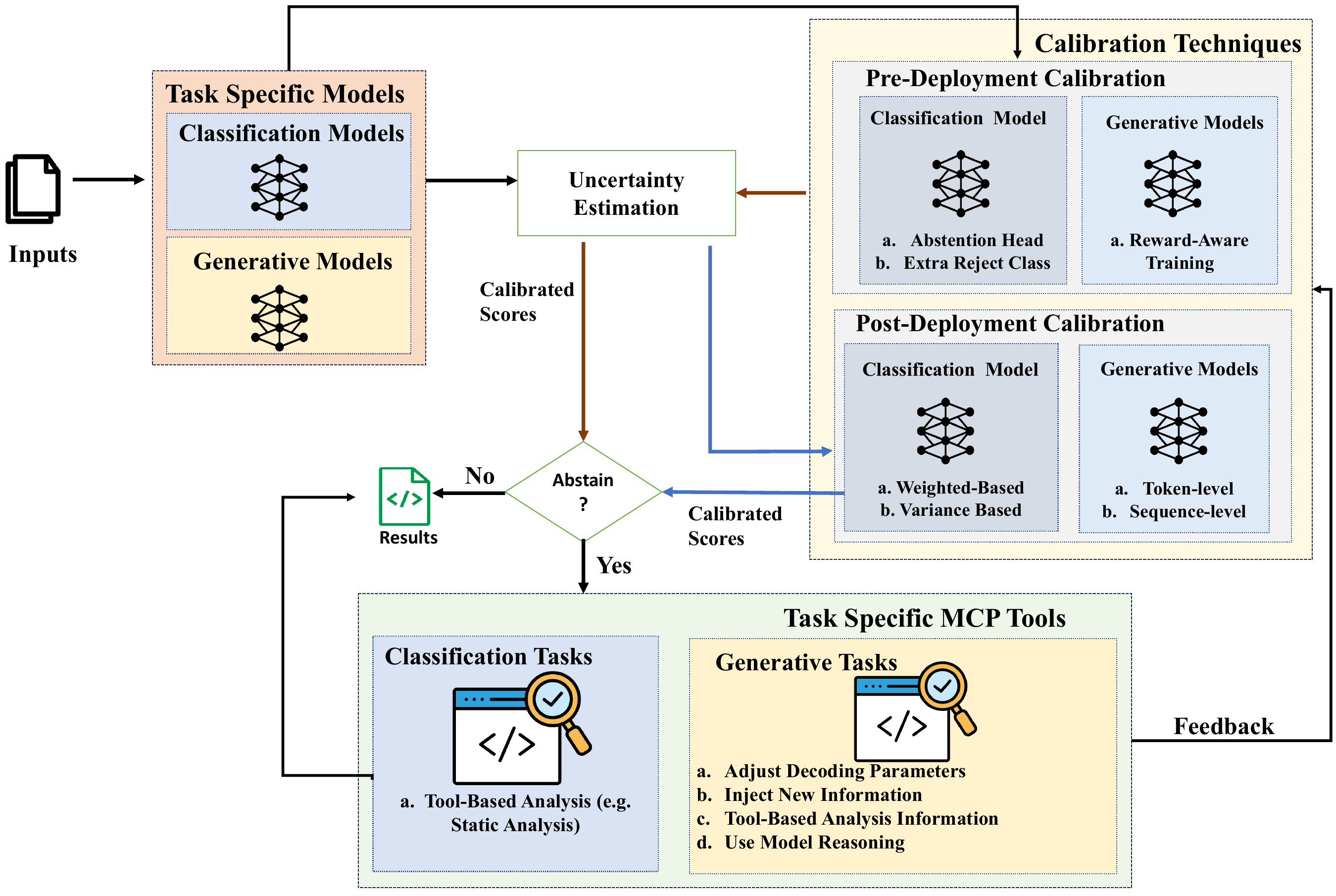}
\caption{Overview of the Proposed Framework} 
\label{fig:Overview}
\end{figure}

We propose a structured framework for deploying code intelligence systems that unifies model uncertainty, selective abstention, and tool-informed recovery into a single decision process (Figure~\ref{fig:Overview}). This design allows models to generate confidence-aware outputs, withhold unreliable predictions, and route uncertain cases through task-specific post-processing. Our goal is to move beyond passive confidence reporting and toward operational systems that act cautiously under uncertainty and adapt their behavior based on calibrated, interpretable signals. 

\textbf{Inputs and model scope.} The framework begins with task-specific inputs, each routed to either a classification model or a generative model~\cite{li2023starcoder,roziere2023code,guo2024deepseek,hui2024qwen2}. Classification tasks include vulnerability detection, defect prediction, or API misuse identification, where the model maps a code representation to a discrete label. Generative tasks include code synthesis, completion, or repair, where the model produces a token sequence given a partial specification or prompt~\cite{tian2023fly, naturalattack, Zhang2023Challenging, lu2021codexglue, yefet2020adversarial,hu2023codes,li2021estimating,van2020tailoring,peng2018t}. The inputs may consist of code snippets, function signatures, or contextual metadata, and are tokenized and encoded according to model requirements. Each model then produces an output, which is a prediction or a completion, along with internal signals used to compute uncertainty.

\textbf{Uncertainty estimation.} The framework begins by extracting internal uncertainty signals from the model, serving as indicators of prediction confidence. Unlike evaluation-centric confidence measures, these signals are used to inform downstream decisions, specifically, whether to trust a model output or defer it.

To support this, we adopt uncertainty estimation strategies that reflect model reliability in both classification and generative tasks. These include metrics derived from predictive distributions as well as methods that quantify model variability under controlled perturbations~\cite{hendrycks2018baseline,steinhardt2016unsupervised,shannon1948mathematical,monarch2021human,gal2016dropout,lakshminarayanan2017simple,vashurin2024benchmarking,rathnasuriya2025codeimprove,rathnasuriya2025framework,rathnasuriya2025fly}. The motivation is to enable the system to make task-specific deferral decisions without relying on costly external validation. These signals are not used directly to filter outputs but instead provide the basis for computing correctness likelihood via calibration.

\textbf{Pre-deployment calibration.} Standard code models are not trained to produce abstention behavior or calibrated uncertainty. Their objectives focus on maximizing prediction accuracy without accounting for decision risk. To enable confidence-aware abstention, we introduce pre-deployment calibration mechanisms that adjust model behavior during training.

For classification models, we introduce mechanisms that allow the model to explicitly abstain. One approach adds a dedicated abstention output pathway trained to identify low-confidence examples. This \textbf{abstention head} is supervised with selective risk-aware losses, encouraging the model to defer when it cannot produce reliable predictions. Another approach extends the label space to include a \textbf{reject class}, allowing the model to route ambiguous inputs away from fixed categories. These strategies convert the model from a closed-world classifier into a system that can explicitly indicate when it should not decide.

For generative models, pre-deployment calibration is implemented via \textbf{reward-aware training}. The model is optimized not only to produce syntactically fluent outputs, but to align generation with correctness proxies such as validation pass rates or task-specific behavior. This changes the behavioral geometry of the model where the model learns to be conservative when uncertain and to structure its generations around downstream acceptability. These techniques require retraining and modify the predictive behavior itself, enabling the model to internalize abstention as a learned action.

\textbf{Post-deployment calibration.}  While pre-deployment techniques change model behavior directly, post-deployment calibration enables abstention over existing models without retraining. The objective here is to convert raw uncertainty signals into calibrated correctness scores that can support abstention decisions.

For classification models, we introduce two forms of post deployment calibration. The first is a \textbf{weighted calibration} scheme, where standard calibration mappings~\cite{guo2017calibration, zhou2024calibration,spiess2024calibration} are trained using loss functions that emphasize certain examples—such as high-error or underrepresented classes—more predominantly. This reflects the real-world need to prioritize reliability over uniform calibration error. The second approach is a \textbf{variance-aware calibration} method, where confidence scores are derived from repeated perturbations of model predictions, and their dispersion is mapped to correctness probabilities. This captures distributional uncertainty more effectively and allows calibrated scores to reflect stability, not just average alignment.

For generative models, post-deployment calibration can operate at multiple granularities. At the \textbf{token level}, uncertainty scores associated with each step of the generation are recalibrated to reflect empirical correctness likelihoods. This is useful in settings like code completion or repair, where partial outputs must be trusted incrementally. At the \textbf{sequence level}, the entire output is assigned a correctness score based on features such as output consistency, structure, and coverage, calibrated against reference data. This holistic calibration allows the system to determine whether the generation, taken as a whole, should be served or deferred. These calibrated scores are then compared to abstention thresholds to decide whether to accept or withhold predictions.

\textbf{Abstention decisions and recovery.} Once a calibrated correctness score is available, the system enforces a policy such that, if the score exceeds a tunable threshold, the prediction is accepted; otherwise, the input is abstained. However, abstention is not terminal. The system activates recovery modules based on the nature of the input and task. For \textbf{generative tasks}, we introduce a \textbf{MCP} layer that resolves abstained predictions via targeted interventions. This layer addresses three distinct failure modes. 

First, when the model fails due to missing task information, such as under-specified prompts, the MCP layer augments the input. It may invoke model reasoning strategies to generate intermediate plans or sketches that clarify intent. If the uncertainty indicates multiple plausible completions, decoding is diversified, and candidates are re-ranked based on follow-up uncertainty estimation. This turns an ambiguous prompt into a controlled exploration.

Second, when failure arises from missing external knowledge, such as unknown libraries or undocumented behaviors, the system injects contextual information, such as documentation snippets or relevant specifications. In addition, lightweight analyzers or compilers may be invoked to validate outputs. These tools convert knowledge uncertainty into evidence the model can act on, reducing epistemic uncertainty and improving downstream calibration.

Third, if the model faces a capability gap where the input falls outside the model’s learned competence, the MCP system enforces simplifications. This may include constraining generation length, enforcing intermediate validation steps, or breaking the task into smaller subtasks. If uncertainty remains high after such interventions, the system abstains definitively, ensuring correctness is never sacrificed for coverage.

For \textbf{classification tasks}, abstention triggers domain-specific recovery. Static analyzers, program slicing tools, or rule-based validators are applied to the input. These tools assess properties such as control flow, data dependencies, or security patterns to validate or reject predictions. This external analysis layer ensures that classification decisions are always supported by either model confidence or symbolic evidence. This work reframes abstention as a principled decision-making mechanism rather than a fixed thresholding heuristic in future code intelligence systems.

\section{Contributions}
In summary, this research makes the following contributions:

\begin{itemize}[leftmargin=*,labelsep=0.5em]
    \item \textbf{Unified framework for abstention across code modeling tasks:}
We envision a framework that generalizes across classification and generative tasks, enabling selective abstention and structured recovery under uncertainty.

    \item \textbf{Task- and model-specific uncertainty calibration:}
We introduce calibration strategies tailored to both classification and generative models.

\item \textbf{Pre- and post-deployment abstention mechanisms:}
We demonstrate how models can be trained to abstain explicitly with abstention heads, reject classes or retrofitted via calibrated confidence thresholds without retraining.

\item \textbf{MCP-guided abstention recovery:} We define a MCP policy layer for resolving abstained predictions via tool-guided interventions. 

\item \textbf{Abstention as an actionable decision primitive:}
We reconceptualize abstention not as failure, but as a route to recovery. The framework incorporates fallback execution and optional feedback loops that adapt to uncertainty in real time.

\item \textbf{Open science and reproducibility commitment:}
All code, configuration files, and calibration datasets will be released under a permissive license upon acceptance. 

\end{itemize}

\section{Preliminary Results}
\label{study}







We investigate whether uncertainty metrics and calibration techniques developed for language and vision tasks can transfer effectively to pretrained code models under both classification and generation settings. Our evaluation is guided by three core questions: (1) Do uncertainty metrics align with correctness in code tasks? (2) Can calibration techniques produce trustworthy abstention decisions? (3) Does abstention improve accuracy under coverage constraints? The findings below support the necessity of behavioral calibration and abstention-aware deployment.

\subsection{Impact of Weighted Scaling for Generative Models}

We implement a lightweight calibration method for generative models based on a weighted logistic scaling approach. The technique maps raw model confidence scores to calibrated correctness probabilities using a monotonic transformation, where the loss function upweights incorrect predictions to penalize overconfident errors. Due to space constraints, we report results on two representative models: DeepSeek-Coder-7b and CodeLlama-7b on the MBPP+~\cite{dong2025codescore} benchmark. However, we have applied the technique across multiple models and tasks with similar trends. For comparison, we evaluate this against platt scaling and isotonic regression. Results are summarized in Table~\ref{tab:mbpp-calibration}. Our findings show that weighted scaling consistently improves both calibration error (ECE)~\cite{guo2017calibration} and brier score (Brier)~\cite{guo2017calibration} across models. Selective prediction at 80\% coverage achieves over 70\% accuracy, indicating that calibrated abstention decisions help isolate reliable outputs. These results validate our framework’s emphasis on conservative, task-specific calibration mechanisms for generative code modeling.

\subsection{Effectiveness of Logit-Based Calibration in Classification}

We propose a behavioral calibration method that operates directly on model logits, bypassing softmax compression and preserving richer information about class separability. Instead of mapping post-softmax scores, our approach trains a correctness estimator directly on raw logit features using a lightweight regressor. This design allows the model to retain decision boundary information and improve per-instance reliability. We apply this method to defect prediction~\cite{phan2017conv} using DeepSeek-Coder-7b and Qwen-Coder-7b, with results shown in Table~\ref{tab:defect-calibration}. Due to space constraints, only these two models are reported here, but we have evaluated the method across multiple model-task pairs and observe consistent improvements in calibration and selective prediction. The method yields the lowest brier score and ECE scores among tested strategies, and at 80\% coverage, achieves over 90\% selective accuracy. These results highlight the benefit of behavioral calibration, thus aligning confidence to correctness at the instance level, rather than relying solely on distributional metrics.

\begin{table}[t]
\centering
\caption{Calibration Results on MBPP+ ($\downarrow$)}
\label{tab:mbpp-calibration}
\resizebox{\linewidth}{!}{
\begin{tabular}{lcccc}
\toprule
& \multicolumn{2}{c}{DeepSeek\textendash Coder\textendash 7B} & \multicolumn{2}{c}{CodeLlama\textendash 7B} \\
\cmidrule(lr){2-3}\cmidrule(lr){4-5}
Method & Brier & ECE & Brier & ECE \\
\midrule
Base Model                & 0.273 & 0.223 & 0.220 & 0.108 \\
Platt Scaling             & 0.224 & 0.103 & 0.248 & 0.062 \\
Isotonic Regression       & 0.216 & 0.143 & 0.215 & 0.054 \\
Weighted Platt Calibration& 0.162 & 0.072 & 0.172 & 0.045 \\
\bottomrule
\end{tabular}
}
\end{table}

\begin{table}[t]
\centering
\caption{Calibration Results on defect prediction  ($\downarrow$).}
\label{tab:defect-calibration}
\resizebox{\linewidth}{!}{
\begin{tabular}{lcccc}
\toprule
& \multicolumn{2}{c}{DeepSeek\textendash Coder\textendash 7B} & \multicolumn{2}{c}{Qwen\textendash Coder-7B} \\
\cmidrule(lr){2-3}\cmidrule(lr){4-5}
Method & Brier & ECE & Brier & ECE \\
\midrule
Base               & 0.130 & 0.029 & 0.137 & 0.023 \\
Temperature Scaling& 0.131 & 0.035 & 0.137 & 0.023 \\
Platt Scaling      & 0.134 & 0.056 & 0.140 & 0.050 \\
Isotonic Regression& 0.130 & 0.015 & 0.137 & 0.016 \\
Confidence         & 0.098 & 0.012 & 0.089 & 0.011 \\
\bottomrule
\end{tabular}
}
\end{table}

\subsection{Evaluation of Uncertainty Metrics Across Tasks}

We evaluate 16 uncertainty metrics~\cite{hendrycks2018baseline,steinhardt2016unsupervised,shannon1948mathematical,monarch2021human,gal2016dropout,lakshminarayanan2017simple,rathnasuriya2025codeimprove,rathnasuriya2025framework,rathnasuriya2025fly} across multiple tasks: defect prediction, vulnerability detection, and code generation. These include entropy-based scores, confidence margins, variance estimators, and sampling-based disagreement measures. Our findings show that while many metrics correlate with prediction correctness, few can be directly used to implement abstention in a reliable way.

Without calibration, uncertainty metrics often overestimate reliability on erroneous predictions. Even after applying standard calibration mappings, improvements in selective prediction vary widely by task and model. In particular, no single uncertainty metric emerges as a task-agnostic solution. This supports one of our key insights: effective abstention in code tasks requires behavioral calibration, i.e., alignment of internal confidence with correctness at the per-instance level.
Unlike settings where probabilistic confidence reflects quality (e.g., classification on natural images), code models often emit confident but incorrect predictions due to superficial pattern matching, semantic ambiguity, and structural repetition. These characteristics limit the utility of uncalibrated uncertainty metrics and motivate our call for abstention-aware calibration pipelines.



\section{Future Plan}
\label{discussion}

Our current efforts have focused on a systematic investigation of existing calibration methods, uncertainty metrics, and their impact on confidence estimation in pretrained code models. Based on these insights, we have begun building a structured foundation for abstention-aware deployment in both classification and generation tasks. Looking forward, our research roadmap includes both short-term objectives and longer-term ambitions to operationalize, scale, and generalize the proposed framework.

\textbf{Short-Term Plan. } Our short-term goals focus on solidifying the empirical foundation and translating the proposed design into implementable modules: (1) \textit{Publishing empirical findings:} We aim to disseminate our results on the limitations of existing calibration methods and the effectiveness of abstention-aware strategies. These findings will help motivate the community to re-express confidence as a decision-making primitive; (2) \textit{Designing and validating envisioned calibration techniques:} We will systematically implement the techniques described in our envisioned framework including both pre-deployment and post-deployment calibration methods for classification and generative models. This includes reward-aware training, weighted scaling, and logit-based behavioral calibration; and (3) \textit{Integrating MCP tools:} We will implement concrete MCP modules for both model types. For classification, this includes rule-based validators and static analysis backends and for generation, we will integrate constraint-based decoding, prompt augmentation, and lightweight synthesis validators. These milestones are planned for completion within a 12 month timeframe and will result in a reproducible, modular codebase covering the full  pipeline.

\textbf{Long-Term Plan.}  Beyond initial implementation, we aim to generalize, scale, and evaluate the framework in real-world deployment contexts. Our long-term vision includes: (1) \textit{Expanding across tasks, models, and domains:} We plan to extend the evaluation across broader model families and additional code intelligence tasks, such as code summarization, clone detection, and test generation. This will help assess task- and model-specific calibration needs; \textit{Black-box model integration and robustness testing:} We aim to validate abstention strategies in scenarios where model internals are not accessible. This will test the applicability of our techniques under limited observability and motivate proxy calibration methods; (3) \textit{Developer-facing deployment integration:} We will integrate our abstention framework into real-world development tools, including IDEs and CI systems. This will allow us to evaluate not only model behavior, but also the usability and trustworthiness of abstention as perceived by developers; (4) \textit{End-to-end system engineering and next-generation MCP tooling:} We plan to build a full abstention-aware inference stack, supporting calibration-aware serving, feedback-driven refinement, and dynamic tool invocation. Over time, we will co-develop next-generation MCP tools that combine program analysis with model introspection to guide abstention and recovery adaptively; and (5) \textit{Open ecosystem and community contribution:} All tools, datasets, and evaluation protocols will be released with support for easy extension, benchmarking, and reproducibility. This will facilitate downstream research on abstention-aware learning, calibration, and human-in-the-loop code AI. These milestones are planned for completion within a 12 to 18 month timeframe. Together, these directions aim to transform abstention from an evaluation artifact into a principled mechanism for safe, adaptive, and trustworthy deployment of code intelligence models.

\section{ACKNOWLEDGMENTS}
This work was partially supported by NSF grants NSF CCF2146443 and Amazon Trust AI Research Award.

\clearpage

\bibliographystyle{ACM-Reference-Format}
\bibliography{main}
\end{document}